# Imaging of bandtail states in silicon heterojunction solar cells

M. Y. Teferi[1], H. Malissa[1], A. B. Morales-Vilches[2], C. T. Trinh[3], L. Korte[3], B. Stannowski[2], C. C. Williams[1], C. Boehme[1⊤], K. Lips[4,5⊥]

[1]*Department of Physics and Astronomy,
115S, 1400E, University of Utah, Salt Lake City, Utah 84112-0830, USA*
[2]*PVcomB, Helmholtz-Zentrum Berlin für Materialien und Energie (HZB), Schwarzschildstr. 3, 12489 Berlin, Germany*
[3]*Institute for Silicon Photovoltaics, Helmholtz Zentrum Berlin für Materialien und Energie GmbH (HZB), Kekulesstr.5, 12489 Berlin, Germany*
[4]*Berlin Joint EPR Lab (BeJEL) and EPR4Energy, Department ASPIN, Helmholtz-Zentrum Berlin für Materialen und Energie GmbH (HZB), Albert-Einstein-Str. 15, 12489 Berlin, Germany*
[5]*Berlin Joint EPR Lab (BeJEL) Department of Physics, Freie Universität Berlin, Arnimallee 14, 14195 Berlin, Germany*

⊤ email: boehme@physics.utah.edu        ⊥ email: lips@helmholtz-berlin.de

**Abstract**

Silicon heterojunction (SHJ) solar cells represent a promising technological approach towards higher photovoltaics efficiencies and lower fabrication cost. While the device physics of SHJ solar cells have been studied extensively in the past, the ways in which nanoscopic electronic processes such as charge-carrier generation, recombination, trapping, and percolation affect SHJ device properties macroscopically have yet to be fully understood. We report the study of atomic scale current percolation at state-of-the-art a-Si:H/c-Si heterojunction solar cells under ambient operating conditions, revealing the profound complexity of electronic SHJ interface processes. Using conduction atomic force microscopy (cAFM), it is shown that the macroscopic current-voltage characteristics of SHJ solar cells is governed by the average of local nanometer-sized percolation pathways associated with bandtail states of the doped a-Si:H selective contact leading to above bandgap open circuit voltages ($V_{\text{OC}}$) as high as 1.2 V ($V_{\text{OC}} > e \cdot E_{\text{gap}}^{\text{Si}}$). This is not in violation of photovoltaic device physics but a consequence of the nature of nanometer-scale charge percolation pathways which originate from trap-assisted tunneling causing dark leakage current. We show that the broad distribution of local photovoltage is a direct consequence of randomly trapped charges at a-Si:H dangling bond defects which lead to strong local potential fluctuations and induce random telegraph noise of the dark current.

**Introduction**

The theoretical limit of the conversion efficiency of single junction solar cells is well understood and described by the Shockley-Queisser limit (SQL)[1–3]. A consequence of the SQL is that the open circuit voltage ($V_{OC}$) of single junction solar cells are always smaller than the bandgap energy ($V_{OC} < e \cdot E_{\text{gap}}$) and is determined by the quasi fermi splitting $\Delta = E_{fn} - E_{fp}$ that can be achieved at the collecting contacts. $\Delta$ is limited by radiative recombination and well understood in terms of thermodynamics[4–10] and determined by the quasi Fermi level distribution in the device. There have been a few observations of $V_{OC}$ being higher than the bandgap energy in various material systems but it seems that these effects could be explained by series connection of several intrinsic junctions. On the other hand, potential fluctuations that may develop at the selective contacts and in the bulk through charge trapping at bandtail states generated through local disorder[11] is known to largely modulate the potential landscape of solar cells and generally lead to a decrease of $V_{OC}$. However, on the atomic scale it is possible that local enhancements of $V_{OC}$ can happen that result in above bandgap $V_{OC}$ as is indicated by the sketch in Fig. 1. Such effects will not be observable with conventional contacts since they would average laterally over the local fluctuations. In this report we will show that with nanometer conductive probes as provided by conduction AFM (cAFM) we are able to resolve such effects in conventional crystalline silicon (c-Si) heterojunction (SHJ) solar cells with hydrogenated amorphous silicon (a-Si:H) as selective contact layers[12–14]. Different from previous cAFM on silicon[15,16] we probe the dark and photocurrents that flow vertically through the selective contact. With this technique, we are able to detect local IV curves with $V_{OC}$ at RT which are as large as 1.2 V. We show that such large $V_{OC}$ are in accordance with thermodynamics and are the result of charge transfer and injection through a-Si:H bandtail- and defect-states at the charge selective SHJ interface[17–21] which leads to stochastic charge trapping and hence to fluctuations of the device performance which is observed as random telegraph noise (RTN) with up to 100% modulation. This is orders of magnitude larger than what has been reported before for doped a-Si:H[22–25]. From the analysis of the cAFM images we can reconstruct a density of states which suggests that we image bandtail states in a-Si:H with a minimum localization of about 1 nm. We suggest that these states are induced by density fluctuations in the amorphous network that have recently been reported about[26].

**Experimental Approach**

We conducted a study of charge percolation across SHJs using ultra-high vacuum (UHV) conduction atomic force microscopy (cAFM) where the a-Si:H surface of the SHJ solar cell was scanned by a platinum AFM tip. For cAFM, tip-sample interaction force is used to gain topographic surface information and use this as feedback observable for the position control of the probe tip, while the tunneling current between the sample and probe tip is measured, allowing for very sensitive local surface conductivity measurements with the lower Å-resolution[27–30]. cAFM has been used successfully to analyze depth-dependent currents in perovskite solar cells[31] or the role of structural defects through conductive-tomographic AFM in CdTe solar cells[32,33]. With cAFM it was also shown that at the a-Si/c-Si interface a 2D conductive layer exists that is responsible for some of the remarkable feature of the SHJ solar cell[15]. However, in none of those cAFM reports on solar cells current limitations due to localized states could be detected.

Figure 1a displays a sketch of a cAFM setup with a SJH device that was processed identically to previously reported and well characterized macroscopic SHJ solar cells with device efficiencies of >20% and 23.65% (certified) on smooth and textured wafers, respectively[34–36]. The main difference of the structures used for cAFM experiments here, compared to complete solar cells, is that they did not have a transparent conducting oxide (TCO) and metal contact, since the Pt tip assumed this role on a nm-scale during the cAFM experiments. For the back contact, the TCO was in direct contact with the cAFM sample holder. The cAFM setup was based on an Omicron Nanotechnology Oxford Instruments LT STM/AFM system which is equipped with a quartz tuning fork in qPlus configuration[27,30] and a commercially acquired[37] Pt tip. The tuning fork resonance frequency $f_0$ was ~30 kHz while a constant amplitude of ~1 nm was maintained under UHV conditions of ~$10^{-10}$ mbar. The quartz tuning fork allows for cAFM measurements in either darkness or under well-defined illumination conditions, in contrast to optically detected AFM where the detection light can induce spurious photo-effects. Note that all experiments reported here were performed at $T$ = 300 K if not stated otherwise.

The SHJ devices studied were based on a double-sided polished n-type <100> oriented float-zone silicon wafer, with details about the sample stack and its preparation being described in the Methods section below. Before devices were introduced into the UHV chamber of the cAFM setup, the oxide layer and other contaminants on the uncontacted (p)-a-Si:H emitter contact layer were removed by application of a brief HF drop with 1% HF solution to the front surface of the device. The device was then mounted on the Omicron sample holder and brought into UHV conditions within 5 min after the HF dip to assure an oxide-free surface for the non-contact mode cAFM measurement. Light was applied by a heat-filtered tungsten halogen lamp with an intensity of 475 W/m² (~½ sun). Topographic images obtained from the tuning fork frequency shift, $\Delta f$, were recorded, while, simultaneously, the current between probe tip and sample was monitored such that local charge transport and topography of the SHJ can be compared. The measurement of current-voltage (*I-V*) characteristics required well controlled, lower Å-range tip-sample distances (TSD), $d$, in tunneling range of the sample surface, that had to be maintained at a constant value even when electrostatic forces changed, e.g. due to changing bias conditions.

**Current variations on the nanometer scale**

Figure 1b shows a representative current image obtained under white light illumination and a forward bias $V_{\text{bias}} = 1$ V while Figure 1c displays the simultaneously taken topographic image. The topography and the current images resemble little similarity, albeit both were recorded at the same time on the same area. While the topography shows 5 Å peak-to-peak roughness, the current under illumination (CuI[1]) image reveals local current maxima from femtoampere to tens of picoampere with varying shapes and nanometer-range sized maxima as determined in a procedure discussed in the SI[38]. Within these areas, referred to as current patches (CPs) in the following, the current is enhanced in magnitude compared to the current in the more homogeneous areas (HAs), where the current is much smaller in magnitude. From these images, we conclude that the current flow through the thin, boron doped a-Si:H emitter layer is highly inhomogeneous and positive with strongest deviations within the nm range sized CPs, while the HAs displays predominantly negative CuI. As shown in Figure 1d and 1e, line scans taken at identical positions shown in panels Figure 1b and 1c, no correlation between current and surface morphology is discernable and, thus, the origin of the CPs cannot be attributed to the sample topography. Hence, the CPs identified in Figure 1b can be attributed to nm scale conductivity inhomogeneity of the SHJ surface- or subsurface-regions, rather than to surface morphology.

---

[1] The CuI is the sum of the dark current and the photocurrent (PC). Due to the superposition principle, the PC in classical p-n junction theory is independent of voltage and is determined by the light-induced generation rate and the internal collection efficiency.

In contrast to larger CPs, the smallest CPs that we have observed can be fitted by Gaussian distributions and they typically show diameters with a full width at half maximum (FWHM) of ~1 nm as shown in Figure 1d. We therefore conclude that these features in the cAFM measurements are not governed by the resolution of our setup, which is about 5 Å as determined for $P_b$ centers located at the c-Si/SiO$_2$ interfaces[39].

While the lower limit of the observed CP size is about one nm, the analysis of 340 different CPs recorded under identical conditions ($V_{bias}$ = 1 V, TSD = 7 Å, see Figure S1 in the SI for more detail[38]), we obtain a lognormal distribution of the outer CP diameter for larger CPs, peaking at $d_{CP} = 2.7\text{nm} \pm 1.3\text{nm}$. We note that the observation of CPs at the a-Si/c-Si heterointerface reproduces for different $V_{\text{bias}}$ and is not a specific feature of doping type since we observe similar CPs for SHJ cells with oppositely doped base and emitter ((n, i) a-Si:H/(p)/c-Si) as shown in the Figure S2 of the SI[38]. We therefore arrive at the conclusion that CPs are a common feature of transport through the a-Si:H/c-Si interface.

We have repeated the current mapping experiments represented by Figure 1b on various surface locations and also for different $V_{\text{bias}}$. Figure 2a displays such measurements made between short circuit condition ($V_{\text{bias}}$ = 0 V and $V_{\text{bias}}$ = 1.8 V). The resulting current maps are plotted in Figure 2a obtained with a thoroughly calibrated current detector, assuring that the varying electrostatic force conditions under varying $V_{\text{bias}}$ did not cause changes of the absolute offset of the current detector as conceivable due to potential changes of the TSD or other influences (see the discussion in SI of Figure S3[38]). The data in Figure 2a shows that, under short circuit conditions, we observe a fairly homogenous and mostly constant, albeit not entirely feature-less short circuit current throughout the probed area. With increasing bias, CPs become increasingly pronounced, i.e. the current within the CPs shifts towards the forward direction, while the number of CPs increases. Some CPs appear only within narrow $V_{bias}$ ranges and then disappear towards higher $V_{bias}$. Also, some CPs display a pronounced substructure with more than one local current maximum. Eventually, at larger $V_{bias}$, especially beyond the macroscopic $V_{OC}$, CPs display large positive currents whose magnitudes become exponentially larger, as expected for the forward current of a SHJ solar cell. Remarkably, this macroscopically well-known diodic *I-V* behavior is displayed microscopically only by the CPs, while large HAs continue to show only a negative CuI, even at $V_{bias} = 1.8$ V (see Figure 2c).

**Fingerprint of Urbach tail in nanoscopic PC**

The topography image of the area where the cAFM data sets of Figure 2a were obtained, shows, again, no significant correlation with any of the current images. Thus, in order to investigate the physical nature of the observed CPs, we determined from the current maps the number of CPs per image, i.e. we determined the number of local current extremata by a procedure explained in the SI using *ImageJ* (see Figure S4 in the SI[38]). The plot of the density, $n_{CP}$, of detected CPs per area as a function of $V_{bias}$ shown in Figure 2b reveals a good agreement with an exponential function with $V_{bias}$ from $3 \times 10^{10}$ cm$^{-2}$ to $5 \times 10^{11}$ cm$^{-2}$.

In order to understand the bias dependency of the CP density, we considered the shifts of the quasi Fermi levels of electrons and holes, $E_{fn}$ and $E_{fp}$, respectively for $0.5 < V_{bias} < 1.8$ V. These shifts were obtained from a 2D numerical simulation *TCAD-SENTAURUS*$^{TM40}$ (see Figure S5 in the SI[38]) of the used SHJ structure under 1 sun illumination (with photon absorption in the 8 nm a-Si:H layer being neglected). These calculations showed that $E_{fn}$ shifts nearly linear with $V_{bias}$ in the *p*-a-Si:H region since electrons are minority carriers, while $E_{fp}$ only shifts in the region close to the c-Si interface and is otherwise pinned due to the high doping level. Since the shift of $E_{fp}$ will change the density of occupied bandtail states in the a-Si:H, we can determine the activation energy of the creation of CPs. We find an activation energy, $E_0 \approx 100$ meV, comparable to the Urbach energy associated with the valence bandtails in highly doped a-Si:H[41]. Thus, the increase in CP area density with $V_{\text{bias}}$ corresponds to the change of the density of occupied states around $E_{\text{fp}}$ in the valence bandtail of the (p)-a-Si:H layer. Essentially, this corresponds to the density of valence bandtail states that allow for hole transport, based on the assumption that these are energetically in a range of $kT$ within $E_{\text{fp}}$. We therefore conclude that the observed CPs correspond to valence bandtail states or nm-sized regions with a certain electronic structure where such bandtail states evolve within the (p)-a-Si:H layer. The cAFM current maps essentially represent images of such regions in the thin emitter layer of the SHJ device which have been made visible for the first time to the best of the authors knowledge.

**Open-circuit voltage variation on the nanometer level**

In order to further investigate the nature of the CPs, including their statistical variance, we have measured more than 1000 *I-V* characteristics on various CPs in different, macroscopically well-separated surface locations under identical conditions. In order to scrutinize that the influence of the TSD and, thus, the tip to surface tunneling probability—an experimental parameter—neither affects the measured current or only in a limited and well understood way, these measurements were carried out with randomly varying TSD between 2 Å and 7 Å (see the discussion in the SI[38]). The red plots in Figure 2c display the *I-V* data sets. Remarkably, for the given illumination conditions, all *I-V* characteristics experience about the same short-circuit current, $I_{\text{SC}} = I_{\text{SC}}^{\text{CP}} = 400$ fA, independent of the TSD. This indicates that charge generation is not altering the solar cell's photocurrent (PC) on the length scales of the experimental window (a few micrometer). As it will be shown later, at least throughout the homogeneous sample surface areas, the excess charge carriers (holes) pass through the interface into the hole selective contact without spatial selectivity, i.e. the currents at any measured point within the imaged area, does not just represent the current at the given location but instead, a current that is broadly distributed throughout the area around the probe location. Note that this primary PC as defined by McGlynn[42] is superimposed by the dark current that is only supported through the highly localized CPs and a good proof of the superposition principle of dark and PC.

While the PC appears to be homogeneous, Figure 2c also reveals that both FF and $V_{OC}^{CP}$ vary dramatically for each CP, with $V_{OC}^{CP}$ being even larger in some few cases than the c-Si bandgap ($V_{OC}^{CP} > q \cdot E_{gap}^{Si}$). Figure 2d displays a histogram of the values of $V_{OC}^{CP}$ as obtained from the *I-V* curves in Figure 2c, which exhibits a broad distribution that is centered at $\overline{V_{OC}^{CP}} = 748$ mV $\pm$ 200 mV, which, remarkably, is the macroscopically expected $V_{OC}$ for this type of base material and surface passivation condition, as found from 2D numerical simulation (see Figure S6 in the SI[38]. The electronic characterization of the SHJ solar cell prepared under similar conditions as our test devices studied here, provide $V_{OC}^{exp} = 0.741$ V with $I_{SC}^{exp} = 38.6$ mA/cm$^2$ with an efficiency >23%[34,35]. Thus, the CP data (red) shown in Figure 2c suggest that a macroscopic SHJ solar cell is essentially a parallel circuit of a large number of CPs which represent microscopic SHJs.

As already seen in Figure 2a and 2c, the *I-V* characteristics within the HAs shows nearly voltage independent CuI, with $I_{SC}^{CP} = I_{SC}^{HA}$. In Figure 2c, we have also plotted the average current obtained from HAs (blue dots connected by line). The plotted *I-V* behavior of HAs can be understood by the definition of HAs, which are those areas that do not show CPs within the applied bias range. This definition already implies that *I-V* functions within HAs will reveal no influence of the dark current injected through the CPs and result in values for $V_{OC}^{HA}$ far beyond $V_{OC}^{exp}$. In essence, as long as *I-V* functions are measured at locations where no CPs are seen and, thus, no valence bandtail states are present, the (p,i)-a-Si:H/c-Si interface behaves like a SHJ interface with (*p*)-a-Si:H that allows no charge injection at low biases. Naturally, the $V_{OC}$ of such a macroscopically non-existent interface, can be higher than $V_{OC}^{exp}$ but cannot be larger than $V_{OC}^{Auger\ limit} \approx 0.75$ V for a cell thickness of 200 μm and under AM 1.5 illumination[1].

Since $I_{SC}^{HA}$ is homogeneously distributed, and $I_{SC}^{HA} = I_{SC}^{CP}$, we attribute this lack of spatial selectivity to holes that are injected into the a-Si:H valence band from the highly delocalized two-dimensional electron gas (2DEG) right below the (*p,i*)-a-Si:H/c-Si interface (see Figure 3 and its discussion below). Interestingly, these observations shown in Figure 1 and 2 imply that dark current and PC at SHJ interfaces separate in different percolation regimes: The PC is generated inside the c-Si absorber of the SHJ and the holes as minority carriers diffuse to the 2DHG at the SHJ interface. From here the excess holes percolate anywhere at that interface across the a-Si:H layers (thermally emitted or direct tunneling) through the a-Si:H valence band. On the other hand, the dark current (holes going from the Pt tip to the c-Si) appears to selectively percolate through the valence bandtail state. We further conclude, that the dark current cannot be injected from the Pt tip into HAs. As a consequence, the HAs can be identified as those areas where we directly observe the primary PC of the SHJ device. Their *I-V* characteristics are similar to the PC deduced from numerical simulation (see Figure S6c in the SI[38]). In a macroscopic device with TCO contacts this primary photocurrent cannot be measured and is always superimposed by the dark current that is established through the CPs. As shown clearly from our results, cAFM is capable of spatially separating dark and photocurrent.

The observation of a large distribution of $V_{OC}^{CP}$ is unexpected as $qV_{OC}$ equal to or above the band gap of the absorber is not possible from a macroscopic thermodynamics point of view, taking the band structures, Fermi-energies, and band offsets into consideration. We can exclude that the strong variation of $V_{OC}^{CP}$ is dominated by a TSD artifact, e.g. by the tunneling gap between tip and sample causing a voltage drop or a depletion in one of the a-Si:H layers. Charge simulations shown in Figure S6a and S6b of the SI[38]) that the large distribution of $V_{OC}^{CP}$ around the well-known macroscopic thermodynamic expectation value $\overline{V_{OC}^{CP}}$ is neither related to the tunneling resistance nor to surface depletion. We therefore hypothesize that it is instead due to local nanoscopic potential fluctuations caused by randomly trapped charges. This hypothesis is quantitatively self-consistent as shown by a simulation of local potential fluctuations caused by the Coulomb fields of randomly spread positive and negative point charges which lead to a random distribution of $V_{OC}$ which is expected to show a small TSD dependence as observed experimentally. This result is also consistent with the spread of $V_{OC}^{CP}$ observed in Figure 2d (see also Figure S7 and S8 in the SI[38]). Previous reports have suggested that above bandgap $V_{OC}$ can exist on the nm level if large electric fields (>10 kV/cm) are present as has been observed in ferroelectric photovoltaic devices[4]. With trapped charge in the vicinity of a tail state, electric fields exceeding 500 kV/cm can exist—as compared to the field in the space charge region of the SHJ device being about only 7 keV/cm.

**Random Telegraph Noise (RTN)**.

In order to scrutinize the trapped charge hypothesis described above, we have studied the dynamics of trapping and emission processes in order to see whether this causes dynamical changes to large potential fluctuations resulting in time dependent distributions of *I-V* curves and hence of $V_{OC}$. Trapping times of free charge carriers in shallow bandtail or defect states occur on femtosecond to nanosecond time scales[43–45], too fast for the time resolution of our experimental setup. However, charge reemission to the bands typically takes place on millisecond to second regimes at room temperature for trap depths between 0.7 and 0.8 eV[22,41]. Thus, trapping and emission processes in close proximity of a CP will alter $V_{OC}$ locally and should be detectable in the time dependence of the current measured on a CP.

We have measured the transient CuI at various positions of the sample (both, at CPs and within HAs) as indicated in Figure 2e, which displays a CuI as a function of time obtained at an arbitrarily chosen CP. During the measurement, a constant TSD was maintained, using the force-feedback mode of the AFM. The transient CuI clearly exhibits two discrete current levels between which the current fluctuates. The inset in Figure 2e depicts the histogram of the measured current trace, which clearly shows a bimodal distribution. This confirms that the current transient shown in Figure 2e displays random telegraph noise (RTN) as known from electronic processes which are controlled by randomly charged and discharged electronic states[46].

We have repeatedly measured time-dependent currents on CPs and found RTN in about 25% of the attempts that we have undertaken. No RTN was detected in the HAs. We have then further scrutinized the nature of this effect by verification that the observation of RTN as shown in Figure 2e originates from trapping and reemission of individual electrons in localized electronic states, and not from artifacts, e.g. the presence of impurity molecules (contaminants) on the probe tip or on the surface, as known from previous scanning probe experiments[47,48]. For a detailed experimental analysis we refer to Figure S9 of the SI[38]. From these experiments, we conclude that the observed RTN must be caused by a mechanism that is inherent to the position where the RTN is detected, i.e. due to an electronic trap and, thus, we conclude that the sample current at some CPs is switched between two levels by local charge trapping and reemission. Thus, changes of the local potential due to trapping or reemission of charges can dramatically switch the injection properties around the CPs on and off as it has been reported before for doped a-Si:H macroscopic samples[22–25]. Teuschler et al.[22] determined that the activation energy of charge capture and emission from deep defects in p-type a-Si:H devices was 0.72 eV and 0.87 eV, respectively, for average switching times of a few 100 ms to a few seconds at RT, comparable to what we observe here. Since in our device only about 25% of the CPs show RTN, we can estimate, following the argumentation of Teuschler et al.[22], that the defect density in our doped a-Si:H layer is about one order of magnitude below the density of bandtail states at the given $E_{fp}$, which means $N_{db} \approx 10^{18}$ cm$^{-3}$, well in accordance with reported estimates the *p*-type a-Si:H doping levels used here[41].

In summary, using cAFM experiments on the hole selective a-Si:H contact of state-of-the-art SHJ solar cells at RT, we were capable to identify atomic scale electronic processes which govern the device behaviors. Dark current is governed by trap assisted tunneling through localized states of the valence bandtail states while PC occurs homogeneous across the SHJ area. The tail states are suggested to originate from dense ordered domains that have been reported in a-Si:H[26] wich can cluster into larger regions of up to 10 nm well in accordance with our experimental findings. The localization of the bandtail states of the valence band are around found to be about 1 nm well in agreement with what has been estimated from theoretical simulations[26].

The nanoscopic *I-V* characteristics of SHJ devices is strongly influenced by fluctuating potentials caused by trapped and detrapping of localized charges at defects in the doped a-Si:H layer, producing local potential fluctuations and, thus, local $V_{OC}$ above 0.8 V, in some few cases even higher than the bandgap. This observation does not violate the fundamental device paradigms of SHJ solar cells as neither the bandgap nor the local Fermi level is well defined on the nm level. With a current collecting TCO layer, the observed nanoscopic effects will average out, leading to $V_{OC}$ of 748 mV, in accordance with experiment and macroscopic device simulations.

We also find that the SHJ solar cell at RT provides different pathways for dark current and PC as illustrated by the sketches in Figure 3a through c, which are band diagrams of the SHJ, showing the percolation of PC (a) as well as dark current (b,c). Note that these band diagrams represent electron energies and localization in a single particle picture, except for the localized electronic bandgap states at the interfaces and within the a-Si:H which are represented in a two-particle picture to avoid ambiguity (see the caption of Figure 3).

Our experiments show that the PC can penetrate the surface at any location, independently of whether a-Si:H bandtail states are present (Figure 3a). In contrast, nm-sized CPs emerge under illumination that result from trap-assisted hole injection from the Pt tip through valence bandtail states of the a-Si:H layer into the c-Si where they recombine with electrons injected from the c-Si back contact (Figure 3b). Because of this, the dark current preferentially penetrates the a-Si:H layer stack only at surface locations where bandtail states are present, implying that the performance limiting component of the dark current originates from trap-assisted tunneling in the doped a-Si:H emitter layer of the SHJ device. This effect is observed even for moderate $V_{bias} < V_{OC}$. In order to minimize these tunneling losses, one would need a hole selective a-Si:H layer that does not have bandtail states within reach of the Fermi level of the contact material, since no bandtail states in the energy window of trap assisted tunneling are present at low $V_{bias}$. For device simulation, the effect of local shunting due to tunneling through bandtail states has to be implemented with a physical model of trap-assisted tunneling. The fact that hole trapping at a defect state in close proximity to the bandtail state (Figure 3c) is occurring will lead to a shift of the bandtail state to lower energy thereby charging the tail state negatively. The fact that no energetically resonant free states in the Pt tip are available for extracting the electron (or pushing in a hole from the tip) leads to a complete blockade of the injection. If the hole is emitted again from the defect this will lift the injection blockage leading to a strongly (order of 100%) modulated RTN, in good accordance with our observation.

## Methods

**Sample preparation**: The SHJ samples investigated were prepared at Helmholtz-Zentrum Berlin (HZB) and their structure is shown in Figure 1a. The c-Si wafers used are both, *n*-type and *p*-type, both sides polished Si <100> oriented 4" quarter wafers. After the RCA cleaning, a three minute chemical etching process in 1% diluted hydrofluoric (HF) acid on both sides of the c-Si wafers was performed, followed by drying with $N_2$ gas. The wafers were immediately loaded in the deposition chamber of a two-chamber plasma-enhanced chemical vapor deposition (PECVD) system to grow the *i*-a-Si:H (3 nm) and *p*-a-Si:H (5 nm) layer. Diborane ($B_2H_6$) was used for *p*-type doping. Similarly, *i*-a-Si:H and *n*-a-Si:H layers were deposited on the back side using $PH_3$ as doping source. ZnO:Al was deposited as a back-contact electrode using an in-line DC magnetron sputtering system from Leybold Optics (A600V7). Similar fabrication processes were followed to fabricate (*n*)a-Si:H/(*i*)a-Si:H/(*p*)c-Si/(*i*)a-Si:H/(p)a-Si:H SHJ. More details on the SHJ device fabrication processes can be found elsewhere[49–51]. The SHJ samples were diced (12 mm × 2.85 mm) and clamped onto an Omicron direct-heating molybdenum sample plate holder for cAFM measurements. The native oxide on the front sides of both SHJ samples was removed by a 1% diluted HF etch followed by $N_2$ drying. The samples were then transferred into the load lock of Omicron nanotechnology LT-STM UHV system within less than five minutes under UHV of $10^{-10}$ mBar.

**Probe preparation**: The conductive Pt tips (25Pt300B) were provided by Rocky Mountain Nanotechnology, LCC[37] and were fabricated from solid platinum wires with a tip radius of 20 nm. The tips were bonded by conductive epoxy to the free tune of tuning fork of the qPlus sensor. The sensor with Pt tip was installed in the UHV LT-STM system (Omicron nanotechnology). Tip stability and conductivity were checked by performing STM measurements on 7 × 7 reconstructed

Si (111) surface which was prepared by a standard flash-annealing procedure[27,28,52]. Stable and reproducible atomistically resolved STM images were obtained which verifies the tip stability and conductivity. Once the tip stability and conductivity are verified in this process, cAFM measurements were taken on SHJ samples.

**cAFM measurements**: The free tune of the qPlus sensor oscillates with a nominal resonance frequency of $f_0 \approx 30$ kHz at a constant amplitude of approximately 1 nm. When a constant forward bias voltage is applied and the tip is within tunneling range, topographic and current images can be simultaneously measured. The frequency shift ($\Delta f$) due to tip-sample interaction was kept constant using a height feedback loop. While scanning across the surface, charge carrier percolation paths with a high lateral resolution in the sub-nanometer range can be detected. Additionally, a single point spectroscopy of *I-V* characteristics was feasible. The intensity of the white light illumination was performed with a tungsten halogen lamp that was focused through the window access of the cAFM setup. The light intensity of 475 W/m$^2$ was estimated through comparison with a reference solar cell at the sample position in the cAFM that was calibrated in in a solar simulator (PV measurements QEXL).

## Data Availability

The data sets generated during and/or analyzed in the course of this study can be obtained by contacting the corresponding authors.

## List of abreviations in alphabetical order

AFM             atomic force microscopy

a-Si:H          hydrogenated amorphous silicon

cAFM            conduction atomic force microscopy

CBT             conduction bandtail

CP              current patch

CuI             current under illumination

c-Si            crystalline silicon

Isc             short circuit current

FF              fill factor

HA              homogeneous area

SHJ             silicon heterojunction

PC              photocurrent

UHV             ultra high vaccum

VBT             valence bandtail

Voc             open circuit voltage

RTN             random telegraph noise

RT              room temperature

SI              Supplementary information

TSD             tip-sample distance

2DEG            2-dimensional electron gas

## Acknowledgements

The cAFM measurement reported in this study were supported by the University of Utah Research Foundation award #10049140. KL is indepted to the Deutsche Forschungsgemeinschaft (DFG), which supported the research visits at the University of Utah through the priority program SPP1601. We acknowledge Holger Rhein Tobias Henschel (PVcomB), Eric Amerling, and Laura Flannery for assistance in sample preparation and cleaning and light intensity measurement.

## Author contributions

This study was planned by K.L. and C.B., A.B.M.-V. and B.S. prepared the SHJ structures, M.T. performed the cAFM experiments, M.T., H.M., C.C.W., K.L., C.T.T., L.K., and C.B. participated in the data analysis and discussions. 2D solar cell simulation were performed by C.T.T., L.K. and K.L. M.T., H.M, and C.B. conducted the charge simluations. The manuscript was written by K.L and C.B. with contributions from all authors.

## Materials & Correspondence

Please address correspondence to C. Boehme or K. Lips

# References

1.  Richter, A., Hermle, M. & Glunz, S. W. Reassessment of the limiting efficiency for crystalline silicon solar cells. *IEEE J. Photovoltaics* **3**, 1184–1191 (2013).

2.  George, B. M. *et al.* Atomic structure of interface states in silicon heterojunction solar cells. *Phys. Rev. Lett.* **110**, 136803 (2013).

3.  Shockley, W. & Queisser, H. J. Detailed Balance Limit of Efficiency of p- n Junction Solar Cells. *J. Appl. Phys.* **32**, 510–519 (1961).

4.  Yang, S. Y. *et al.* Above-bandgap voltages from ferroelectric photovoltaic devices. *Nat. Nanotechnol.* **5**, 143–147 (2010).

5.  Limpert, S. *et al*. Single-nanowire, low-bandgap hot carrier solar cells with tunable open-circuit voltage. *Nanotechnology* **28**, 434001 (2017).

6.  Spanier, J. *et al*. Power conversion efficiency exceeding the Shockley-Queisser limit in a ferroelectric insulator. *Nature. Photon.* **10**, 611–616 (2016).

7.  Nechache, R. *et al*. Bandgap tuning of multiferroic oxide solar cells. *Nature. Photon.* **9**, 61–67 (2015).

8.  Yang, B. *et al*. Tuning the Energy Level Offset between Donor and Acceptor with Ferroelectric Dipole Layers for Increased Efficiency in Bilayer Organic Photovoltaic Cells. *Adv. Mater.* **24**, 1455–1460 (2012).

9.  Seidel, J. *et al.* Efficient Photovoltaic Current Generation at Ferroelectric Domain Walls. *Phys. Rev. Lett.* **107**, 126805 (2011).

10. Alexe, M. & Hesse, D. Tip-enhanced photovoltaic effects in bismuth ferrite. *Nat. Commun.* **2**, 256 (2011).

11. Dong, J. & Drabold, D. A. Atomistic Structure of Band-Tail States in Amorphous Silicon. *Phys. Rev. Lett.* **80**, 1928–1931 (1998).

12. van Sark, W., Korte, L. & Roca, F. *Physics and Technology of Amorphous-Crystalline Heterostructure Silicon Solar Cells*. (Springer Berlin Heidelberg, 2012).

13. Stangl, R., Kriegel, M. & Schmidt, M. AFORS-HET, Version 2.2, a Numerical Computer Program for Simulation of Heterojunction Solar Cells and Measurements. in *2006 IEEE 4th World Conference on Photovoltaic Energy Conference* 1350–1353 (IEEE, 2006).

14. Tanaka, M. *et al.* Development of new a-si/c-si heterojunction solar cells: ACJ-HIT (artificially constructed junction- heterojunction with intrinsic thin-layer). *Jpn. J. Appl. Phys.* **31**, 3518–3522 (1992).

15. Kleider, J. P. *et al.* Characterization of silicon heterojunctions for solar cells. *Nanoscale Res. Lett.* **6**, 152 (2011).

16. Wiesendanger, R. *et al*. Hydrogenated amorphous silicon studied by scanning tunneling microscopy. *J. Appl. Phys.* **63**, 4515–4517 (1988).

17. Yoshikawa, K. *et al.* Silicon heterojunction solar cell with interdigitated back contacts for a photoconversion efficiency over 26%. *Nat. Energy* **2**, 1–8 (2017).

18. Masuko, K. *et al.* Achievement of more than 25% conversion efficiency with crystalline silicon heterojunction solar cell. *IEEE J. Photovoltaics* **4**, 1433–1435 (2014).

19. Nakamura, J. *et al.* Development of heterojunction back contact Si solar cells. *IEEE J. Photovoltaics* **4**, 1491–1495 (2014).

20. Smith, D. D. *et al.* Toward the practical limits of silicon solar cells. *IEEE J. Photovoltaics* **4**, 1465–1469 (2014).

21. Yoshikawa, K. *et al.* Exceeding conversion efficiency of 26% by heterojunction interdigitated back contact solar cell with thin film Si technology. *Sol. Energy Mater. Sol. Cells* **173**, 37–42 (2017).

22. Teuschler, T., Hundhausen, M., Ley, L. & Arce, R. Analysis of random telegraph noise in large-area amorphous double-barrier structures. *Phys. Rev. B* **47**, 12687–12695 (1993).

23. Parman, C. E., Israeloff, N. E. & Kakalios, J. Random telegraph-switching noise in coplanar current measurements of amorphous silicon. *Phys. Rev. B* **44**, 8391–8394 (1991).

24. Choi, W. K., Owen, A. E., LeComber, P. G. & Rose, M. J. Exploratory observations of random telegraphic signals and noise in homogeneous hydrogenated amorphous silicon. *J. Appl. Phys.* **68**, 120–123 (1990).

25. Arce, R. & Ley, L. Observation of Random Telegraphic Noise in Large Area a-Si:H/a-Si$_{1-x}$,N$_x$:H Double Barrier Structures. in *MRS Proceedings* vol. 149 675 (Cambridge University Press, 1989).

26. Gericke, E. *et al*. Quantification of nanoscale density fluctuations in hydrogenated amorphous silicon. Preprint at https://arxiv.org/abs/2006.01907 (2020).

27. Giessibl, F. J. The qPlus sensor, a powerful core for the atomic force microscope. *Rev. Sci. Instrum* **90**, 11101 (2019).

28. Ohnesorge, F. & Binnig, G. True atomic resolution by atomic force microscopy through repulsive and attractive forces. *Science* **260**, 1451–1456 (1993).

29. Majzik, Z. *et al.* Simultaneous current, force and dissipation measurements on the Si(111) 7×7 surface with an optimized qplus AFM/STM technique. *Beilstein J. Nanotechnol.* **3**, 249–259 (2012).

30. Giessibl, F. J. Atomic resolution of the silicon (111)-(7x7) surface by atomic force microscopy. *Science* **267**, 68–71 (1995).

31. MacDonald, G. A. *et al.* Methylammonium lead iodide grain boundaries exhibit depth-dependent electrical properties. *Energy Environ. Sci.* **9**, 3642–3649 (2016).

32. Luria, J. *et al.* Charge transport in CdTe solar cells revealed by conductive tomographic atomic force microscopy. *Nat. Energy* **1**, 16150 (2016).

33. Kutes, Y. *et al*. Mapping photovoltaic performance with nanoscale resolution. *Prog.*

*Photovolt: Res. Appl.* **24**, 315–325 (2016).

34. Cruz, A. *et al.* Effect of front TCO on the performance of rear-junction silicon heterojunction solar cells: Insights from simulations and experiments. *Sol. Energy Mater. Sol. Cells* **195**, 339–345 (2019).

35. Wang, E.-C. *et al.* A simple method with analytical model to extract heterojunction solar cell series resistance components and to extract the A-Si:H(i/p) to transparent conductive oxide contact resistivity. in *15th International Conference on Concentrator Photovoltaic Systems (CPV-15)* vol. 2149 040022 (2019).

36. Morales-Vilches, A. B. *et al.* ITO-Free Silicon Heterojunction Solar Cells with ZnO:Al/SiO 2 Front Electrodes Reaching a Conversion Efficiency of 23%. *IEEE J. Photovoltaics* **9**, 34–39 (2019).

37. Rocky Mountain Nanotechnology. Rocky Mountain Nanotechnology, LLC. *https://rmnano.com/probes* https://rmnano.com/probes (2003).

38. 'See supplementary information to this article'.

39. Ambal, K. *et al.* Electrical current through individual pairs of phosphorus donor atoms and silicon dangling bonds. *Sci. Rep.* **6**, 18531 (2016).

40. SYNOPSYS. *Sentaurus Structure Editor User Guide*. (synopsis, Inc, 2011).

41. Street, R. A. *Hydrogenated Amorphous Silicon*. (Cambridge University Press, 1991).

42. McGlynn, S. P. Concepts in Photoconductivity and Allied Problems. *J. Am. Chem. Soc.* **86**, 5707–5707 (1964).

43. Uhd Jepsen, P. *et al.* Ultrafast carrier trapping in microcrystalline silicon observed in optical pump-terahertz probe measurements. *Appl. Phys. Lett.* **79**, 1291–1293 (2001).

44. Nebel, C. E. & Bauer, G. H. Tail-state distribution and trapping probability in a-Si:H investigated by time-of-flight experiments and computer simulations. *Philos. Mag. B* **59**, 463–479 (1989).

45. Fekete, L. *et al.* Ultrafast carrier dynamics in microcrystalline silicon probed by time-resolved terahertz spectroscopy. *Phys. Rev. B* **79**, 115306 (2009).

46. Neri, B., Olivo, P. & Riccò, B. Low-frequency noise in silicon-gate metal-oxide-silicon capacitors before oxide breakdown. *Appl. Phys. Lett.* **51**, 2167–2169 (1987).

47. Chen, F., Huang, Z. & Tao, N. Forming single molecular junctions between indium tin oxide electrodes. *Appl. Phys. Lett.* **91**, 162106 (2007).

48. Xu, B. & Tao, N. J. Measurement of single-molecule resistance by repeated formation of molecular junctions. *Science* **301**, 1221–1223 (2003).

49. Mazzarella, L. *et al.* Comparison of TMB and B2H6 as precursors for emitter doping in high efficiency silicon hetero junction solar cells. in *Energy Procedia* vol. 60 123–128 (2014).

50. Mazzarella, L. *et al.* Nanocrystalline n-type silicon oxide front contacts for silicon

heterojunction solar cells: Photocurrent enhancement on planar and textured substrates. *IEEE J. Photovoltaics* **8**, 70–78 (2018).

51. Mazzarella, L. *et al.* Nanocrystalline silicon emitter optimization for Si-HJ solar cells: Substrate selectivity and CO2 plasma treatment effect. *Phys. Status Solidi Appl. Mater. Sci.* **214**, 1532958 (2017).

52. Chen, C. J. *Introduction to Scanning Tunneling Microscopy*. (Oxford University Press, 2007).

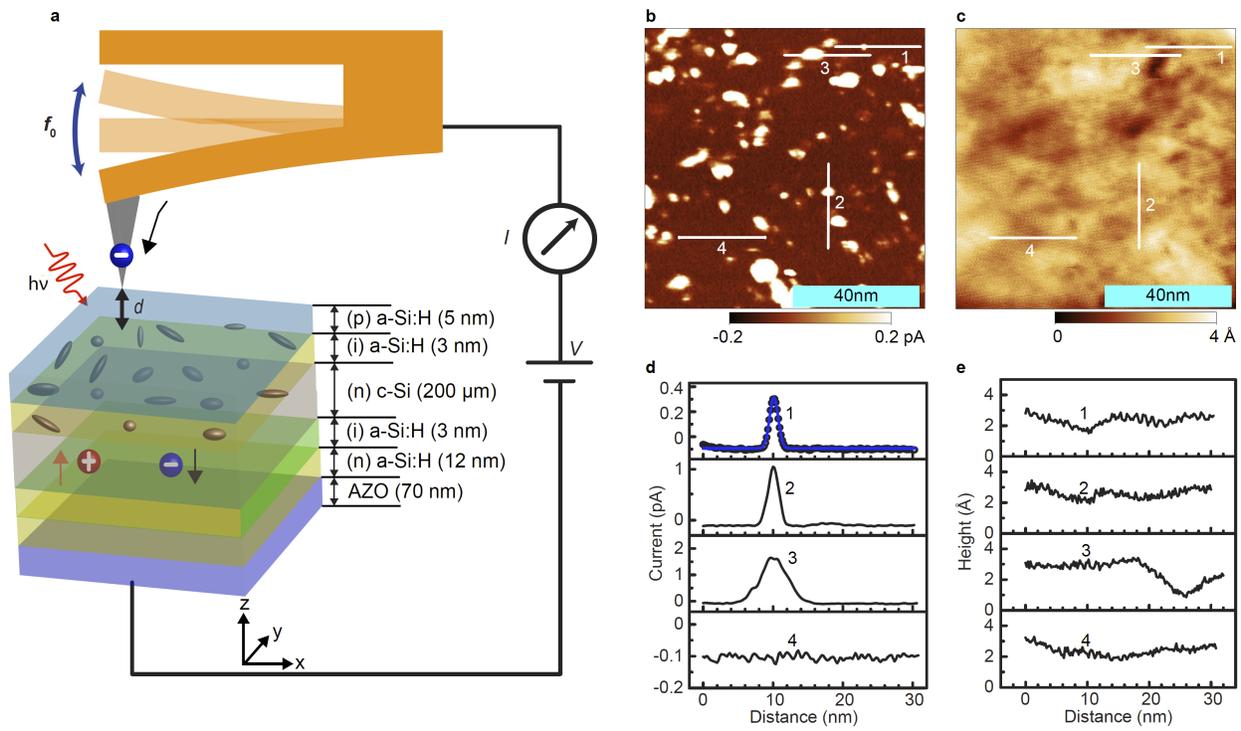

*Figure 1: a) Sketch of the RT UHV cAFM experiment on SHJ solar cell structures* allowing for dark and photoconductivity measurements at various surface locations with atomic scale resolution using a scanning probe, while AFM force feedback allows for control of TSD. White light illumination was achieved with a tungsten halogen lamp. *b) Current under illumination and c) topographic image* of a $100\times100$ nm² scan area measured using cAFM under illumination and forward bias condition on the SHJ shown in a). The topographic and current images are obtained simultaneously. d) Plots of different current profiles along the line cuts 1-4 shown in b). e) The plot of the topographic height profile along the line cuts 1-4 shown in c). The blue curve in d) is the fit to the data with a gaussian profile of FWHM 1 nm.

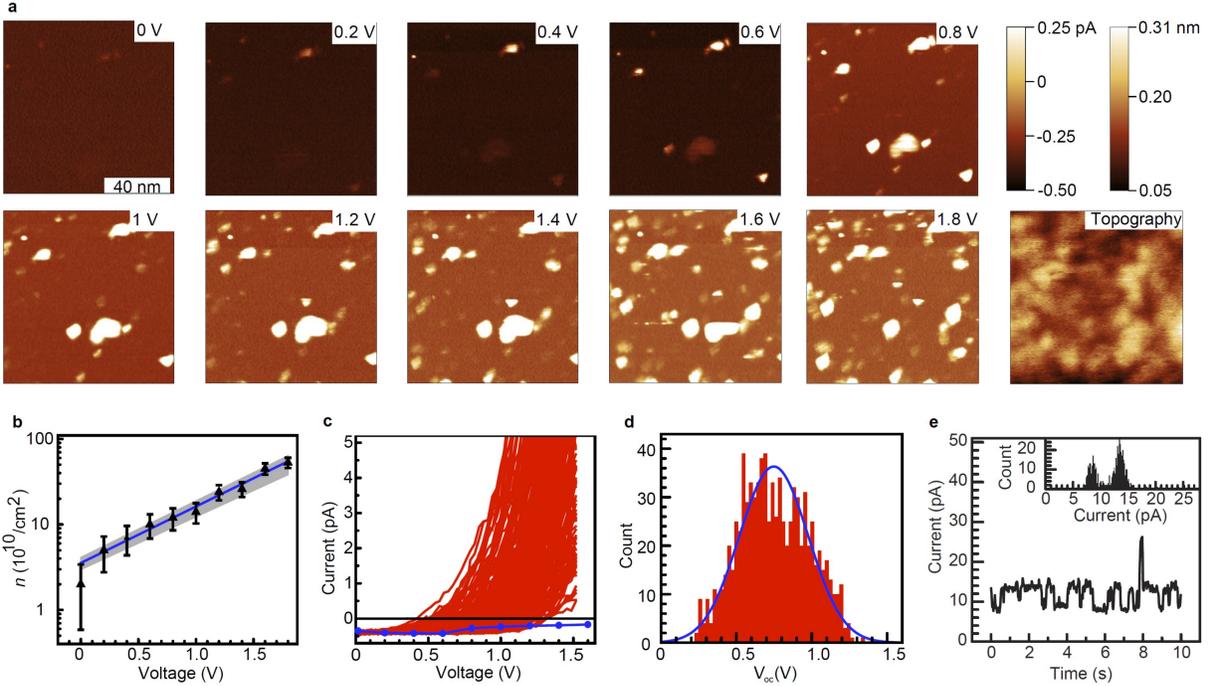

*Figure 2: a) Bias dependence of cAFM current maps of the SHJ sample shown in Fig. 1a.* The panels show RT cAFM current images at indicated $V_{bias}$ measured under illumination at constant TSD. The topography was measured simulatenously and shows no significant correlation with the current maps. Current and topographic scale are shown in color code. *b)* Density of CPs as function of $V_{bias}$ extracted from panels a) through a procedure described in SI [38] with error bars indicated in the figure. *c)* the red curves indicate 1000 I-V curves extracted at various CPs (for details see text) I-V characteristics were obtained under ~0.5 suns illumination at various TSD between 2 Å and 7 Å in non-contact mode while the force-feedback loop was turned off. The blue dotted line represent the IV curves obtained through integration over the HAs of the current maps in a). In contrast to the I-V curves recorded in CPs (red lines), diodic behavior is not apparent. *d)* Histogram of the distribution of the open-circuit voltages $V_{OC}^{CP}$ as extracted from the I-V data displayed in c). The blue curve represents a fit with a normal distribution. *e)* Time evolution of the current through CP measured at $V_{bias} = 1\ V$ exhibiting RTN. The inset shows the current distribution recorded over 10 s showing a pronounced bimodal distribution.

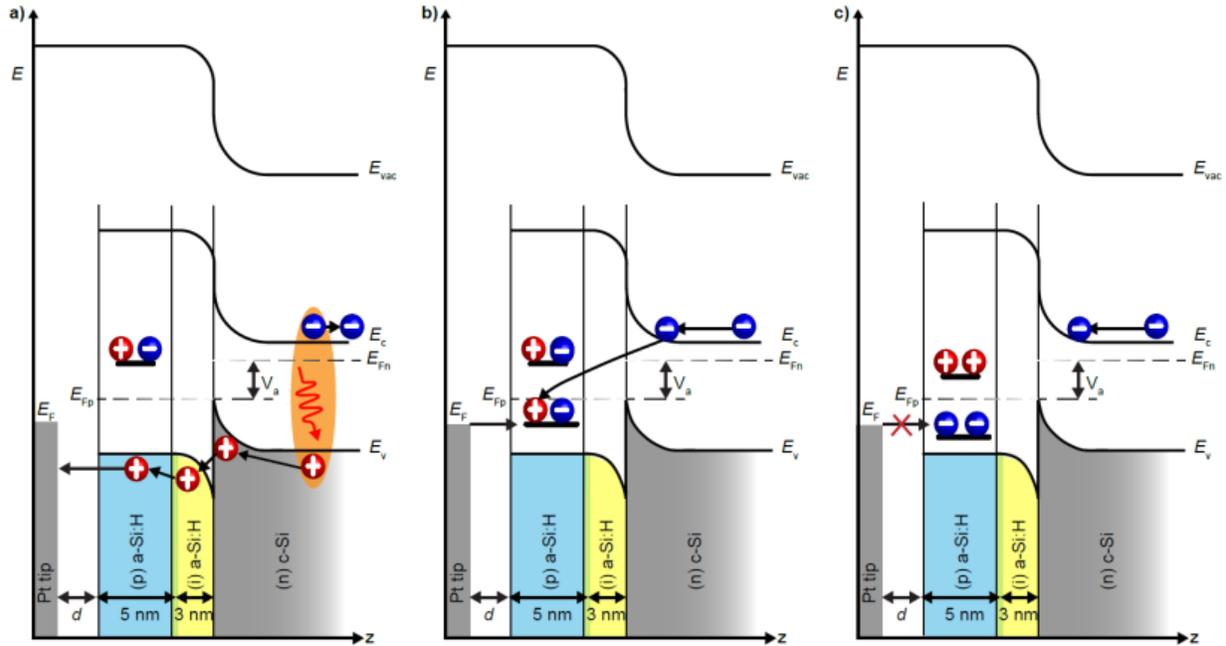

*Figure 3: Illustrations of SHJ charge transfer percolation pathways of photo- and dark-currents at HAs (a) as well as CPs (b, c).* Note that interface band diagrams are usually single electron state illustration and, thus, charge carriers such as electrons in the conduction bands (blue, -) and holes in the valance bands (red, +) are illustrations of individual occupied and unoccupied electron states, respectively. In contrast, in order to avoid ambiguity, the charge states of localized electronic bandgap states such as tail and dangling bond states are represented here in a two-electron picture where (+-),(++), and (--) represent neutral, singly positive as well as singly negative charge states, respectively. The images presents a near open-circuit situation where the quasi Fermi-level splitting $V_a \sim V_{bias} = V_{OC}$. **a)** Charge extraction of photogenerated holes in HAs through tunneling of electrons in the Pt tip (below the Pt Fermi level) into the valence band of a-Si:H. **b)** Injection of holes from the Pt tip in to the SHJ through electron tunneling of electrons out of the a-Si:H valence band-tail state into unoccupied electron states above the Pt Fermi level. This process represents dark current at CPs. **c)** Hole injection blockade at a CP due to the presence of positive charge that is trapped at a nearby db defect state deep within the a-Si:H band gap. The trapped charge causes the band-tail state energy level to shift closer to the valence band such that it becomes negatively charged. This leads to a complete blockade of current injection, and hence, to a significant dark current change at the CP. The dark current will fluctuate and generate RTN, when the db state is randomly charged and discharged repeatedly. The quasi-Fermi levels $E_{fn}$ and $E_{fp}$ are indicated by dashed lines and represent the thermodynamic situation of the band diagram, blue and red arrows indicate transitions of electrons or holes, respectively.

# Supplementary Information

## Imaging of bandtail states in silicon heterojunction solar cells

M. Y. Teferi[1], H. Malissa[1], A. B. Morales-Vilches[2], C. T. Trinh[3], L. Korte[3], B. Stannowski[2], C. C. Williams[1], C. Boehme[1]⊤, K. Lips[4,5]⊥

[1]*Department of Physics and Astronomy,
115S, 1400E, University of Utah, Salt Lake City, Utah 84112-0830, USA*
[2]*PVcomB, Helmholtz-Zentrum Berlin für Materialien und Energie GmbH (HZB), Schwarzschildstr. 3, 12489 Berlin, Germany*
[3]*Institute for Silicon Photovoltaics, Helmholtz Zentrum Berlin für Materialien und Energie GmbH, Kekulesstr.5, 12489 Berlin, Germany*
[4]*Berlin Joint EPR Lab (BeJEL) and EPR4Energy, Department ASPIN, Helmholtz-Zentrum Berlin für Materialen und Energie GmbH (HZB), Albert-Einstein-Str. 15, 12489 Berlin, Germany*
[5]*Berlin Joint EPR Lab (BeJEL) Department of Physics, Freie Universität Berlin, Arnimallee 14, 14195 Berlin, Germany*

⊤email: boehme@physics.utah.edu          ⊥email: lips@helmholtz-berlin.de

1.     **Nano-sized local current maxima distribution**

Figure S1 shows the distribution of the size of local current maxima of current maps for (p) a-Si:H/ (i) a-Si:H/ c-Si heterojunction solar cell. The whole data set was extracted from the local current maxima of eight current maps acquired under nominally the same conditions ($V_{bias} = 1\,V$, $d = 7$ Å, and under white light illumination) from over 340 current patches (CPs). Each data point in Figure S1 was obtained from the full-width half maxima of the Gaussian fit to the outer larger diameter ($d_{CP}$) of each CP (see an example in Figure 2 of the main text). The size distribution follows a lognormal type of distribution with an average diameter of $d_{Cp} = 2.7$ nm $\pm$ 1.3 nm. The minimum $d_{CP}$ is approximately 10 Å. In the past, other groups have reported localization length of bandtail states of 8.5 Å[1], 11 Å[2], 12 Å[3] measured using different techniques by assuming a single exponential envelope of the wave function which falls off isotropically. Our result of the smallest size of CP of 10 Å is in agreement with assumptions in literature, and the smallest CPs that we observe are most likely surface states in the (p) a-Si:H layer may involve single trap assisted tunneling. However, the observed wide size distribution of CPs (see Figure S1) implies that each larger size CP may involve different percolation paths (domains) from c-Si to the p-layer and charge transport in this case involves trap assisted tunneling through two or more valance bandtail states. The size distribution not only provides information about the involvement of two or more valance bandtail states but also may represents depth profile of electronic states. Bandtail states can exist across the volume of the (p) and (i) a-Si:H layer and they can be more accessible to the Pt tip if they are closer to the p-layer surface than deeper states in the volume of both layer. Such type of size variation due to depth profile of electronic states on non-conductive dielectric samples have been observed using a technique called dynamic tunneling force microscopy which is sensitive to single trap state[4,5]. However, we did not observe CP size below 10 Å. Therefore, we argue that larger CPs are due to a cluster of tunneling steps across several band-tail states (domains).

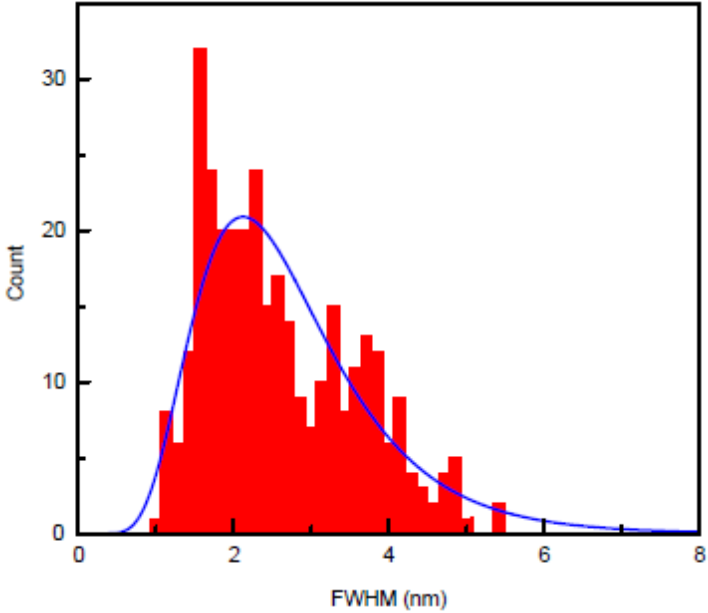

*Figure S1. The Histogram of the full-width half maxima of the size of CPs (outer diameter) for (p) a-Si:H/ (i) a-Si:H/ c-Si HIT solar cell*

2. **Current variation on nanometer-scale on (n,i) a-Si:H/(p) c-Si HIT solar cell**

Figures S2a and S2b show current and topographic images obtained under white light illumination and forward bias of $V_{bias} = 1$ V taken simultaneously on the same area, respectively. As can clearly be seen from the images and their line cut profiles, both, topography and current under illumination (CuI) image show no correlation. While the topography image shows peak-to-peak surface corrugation (or roughness) of about 5 Å, the CuI image reveals local current maxima with varying shapes and nm-range sizes, referred to as current patches (CPs). The CPs have larger or enhanced current as compared to the current in the more homogeneous areas (HAs). Similar type of local conductivity variation on the nanoscale have been reported for quantum dots with defects[6], high-k dielectric oxides[7], perovskite solar cell samples[8], and through phosphorous and dangling bond pairs at the silicon/SiO$_2$ interface[9] under ultra-high vacuum condition. Our results reveal that the current flow through the thin, boron-doped a-Si:H emitter layer is highly inhomogeneous and shows strong variation within the nm-range sized CPs and HAs. Therefore, we attribute these strong nm-range sized CPs and HAs to nm scale conductivity variation of the HIT surface- or subsurface-regions. A similar observation is made for (p,i) a-Si:H/(n) c-Si HIT solar cell as presented and discussed in the main text.

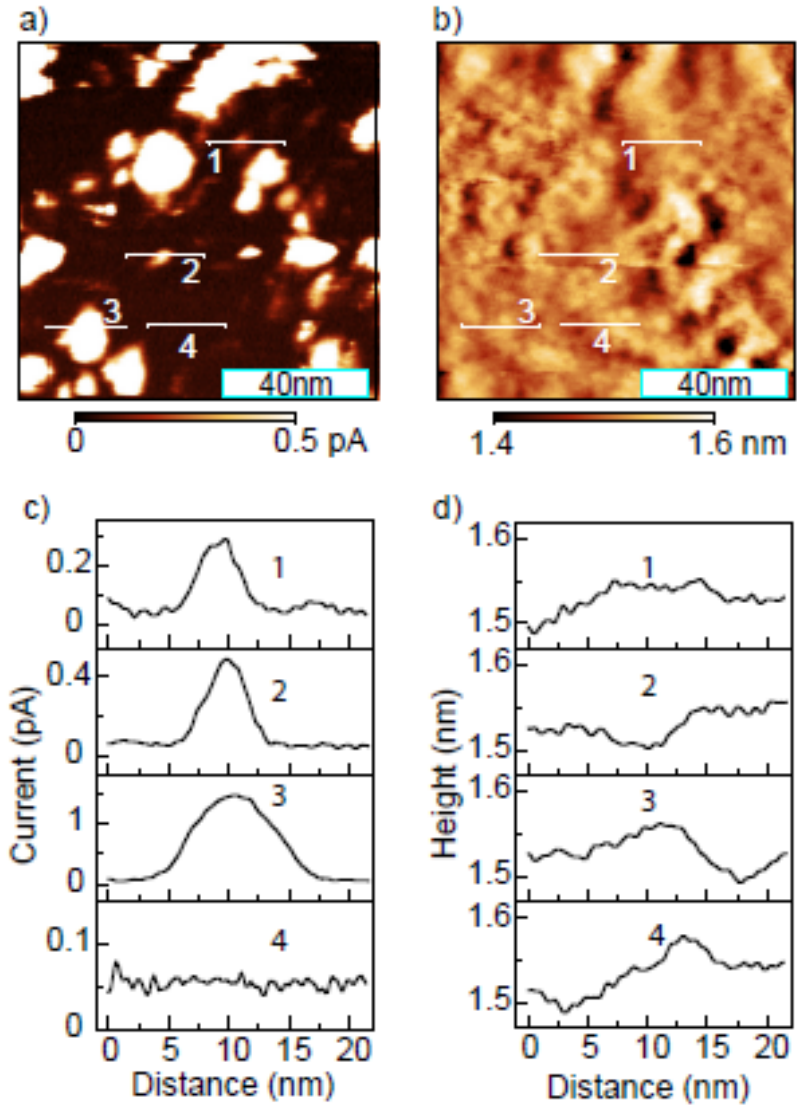

*Figure S2.* a) The current image of a 100×100 nm$^2$ scan area measured using cAFM under white light illumination and forward bias condition $V_{bias} = 1$ V on a (n,i) a-Si: H/(p) c-Si heterojunction sample. b) The topographic image of the same area, as obtained simultaneously with the current image shown in a). c) Plots of different current profiles along the line cut 1, 2, 3, 4 shown in a). d) The plot of the topographic height profile along the line cuts 1, 2, 3,4 shown in b). The comparison of the topography and current data in a) and b) and, thus, in c) and d) show that there is no discernable correlation between current and topographic images.

## 3. Verification of constant offset current of current detector on the gap and applied bias

Figure S3 displays the measured offset current as a function of time for different tip-sample gap distance (TSD) and applied bias voltage, $V_{bias}$. The plots in Figure S3a show the current transient at different tip-sample gap distance where the TSD was maintained constant by using the force feedback control with the respective set frequency shift. All of these plots were acquired at $V_{bias} = 0$ V and under dark condition. These results confirm that the absolute offset current of the current detector remains constant for different TSD. We have also acquired current images at different TSD under the dark condition, and zero bias (not shown here) and the results show that the offset current does not change with TSD. Figure S3b depicts the transient offset current at different applied bias voltage under dark condition when TSD is large which corresponds to a frequency shift $\Delta f = -5$ Hz and the Pt tip is not in tunneling range. These results confirm that the current detector offset current does not depend on $V_{bias}$.

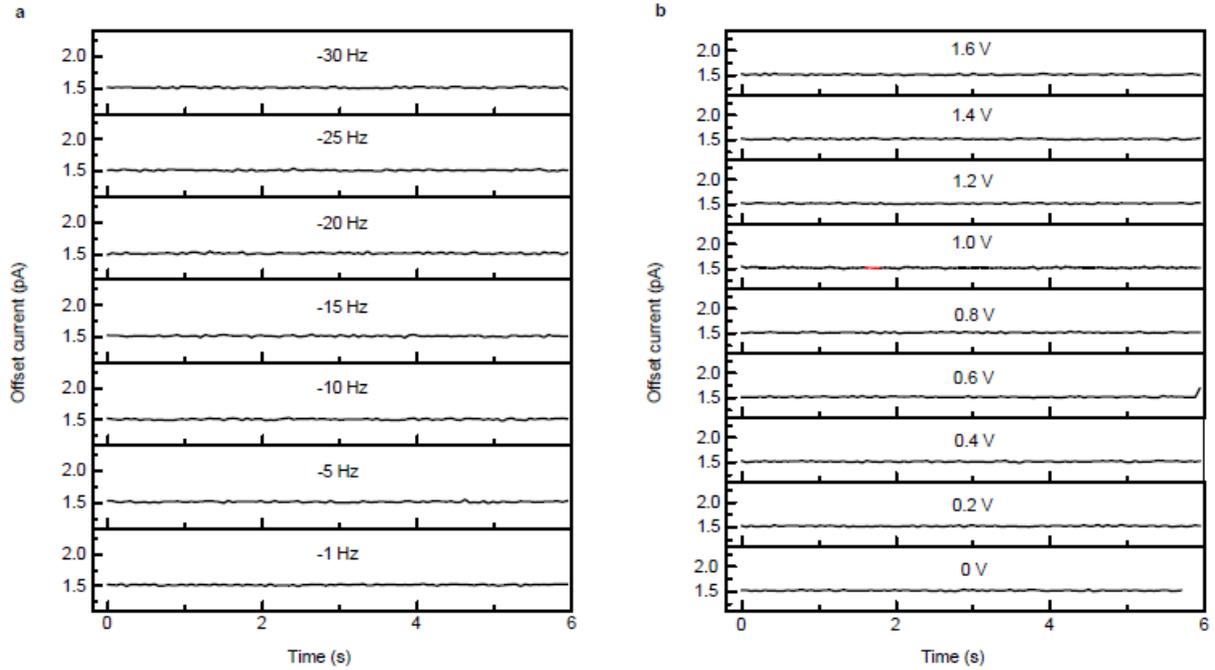

*Figure S3. a) Offset current as a function of time for different TSD which is controlled by the respective frequency shift. b) Offset current as a function of time for different $V_{bias}$. Note that the offset current for panel b is acquired at large TSD.*

## 4. Density of state determination

Figure S4 displays the procedure we followed to determine the density of states from current images acquired at different $V_{bias}$ shown in the main text of Figure 3. Figure S4a shows the current image measured at $V_{bias} = 1.8$ V as displayed in the main text of Figure 3. This original image acquired using cAFM is converted into a binary image using the ImageJ processing software[10,11] with global thresholding which identifies the CPs (black regions) from the background (HA: white regions) as shown in Figure S4b. Common global thresholding was applied for all current images.

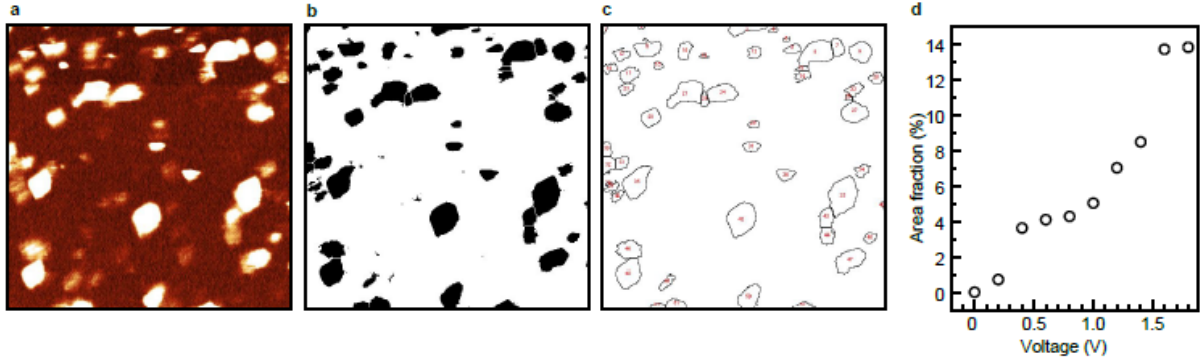

*Figure S4. Determination of density of states. a) current image obtained with $V_{bias} = 1.8\,V$, as shown in Fig. 2 of the main text. See Fig. 2 also for the color to current conversion legend. b) the current image in a binary black and white after an appririate black-to white threshold current has been identified. c) The number of current patches identified through local current maxima (red dots). d) area fraction of the CPs as a function of $V_{bias}$.*

The number of CPs and their total area (the sum of the areas of the individual CPs) are obtained from each current image following this procedure, an example of which is shown in Figure S4c. Figure S4d shows the area fraction (total area of CPs divided by the total area of the current image) as a function of $V_{bias}$. The number of CPs per area for all current images as a function of $V_{bias}$ is shown in the lower right panel of Figure 3 in the main text. This clearly shows that the areal density of states increases with $V_{bias}$ (see detailed explanation in the main text). Moreover, the area fraction of CPs also increases with $V_{bias}$, which indicates an increase of the number of CPs.

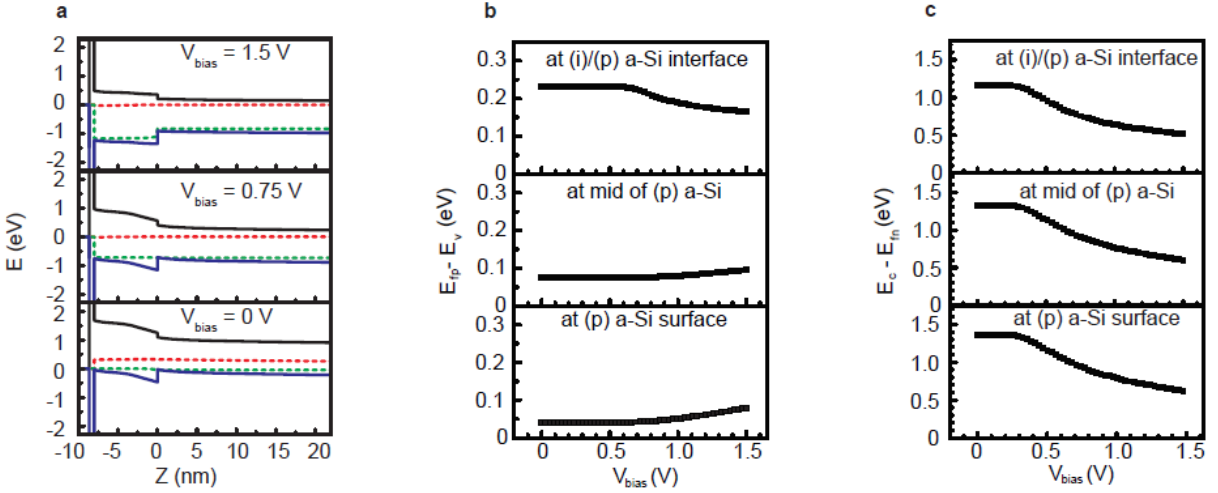

*Figure S5. 2D numerical simulation of energy band diagram under one sun illumination. a) Energy band diagram at TSD of 0.75 nm for applied bias voltage of 0 V, 0.75 V, 1.5 V as indicated in the figure. The black line represents conduction band, the red dash line represents quasi Fermi level of electrons, the green dash line represents quasi Fermi level of holes, and the bluse line represents the valance band. b) The shift of quasi Fermi level of holes with respect to the valance bandas as a function of applied bias at difference location of SHJ as indicated in the figure. c) The shift of quasi Fermi level of electrons with respect to the conduction band as a function of applied bias at difference location of SHJ as indicated in the figure.*

## 5. Two dimensional numerical simulation of the HIT structure studied in the cAFM experiment with SENTAURUS

2D-device simulation was performed using TCAD-SENTAURUS™ [12]. The structure of the simulated device is the same as in Figure 1a in the main text, except the TCO was omitted and silver was used as metallization at the back contact. Under illumination (1 sun), simulated optical generation was obtained using optical beam absorption method, in which photon absorption in c-Si is calculated following Beer's law. Current flow through vacuum gap was governed by barrier tunneling with effective mass in vacuum is taken as 1. Input electrical parameters of a-Si:H films were taken from standard a-Si layers as used in the silicon heterojunction device simulator AFORS-HET[13,14].

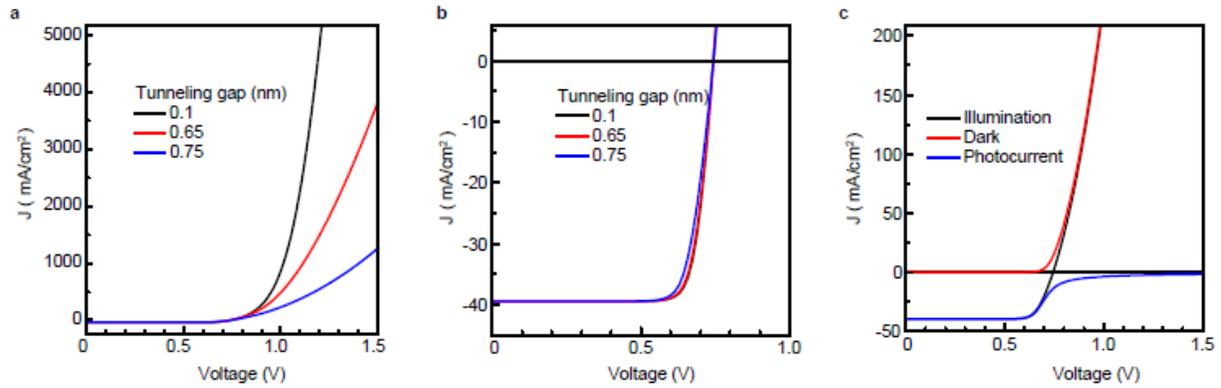

*Figure S6. 2D numerical simulation of the I-V curve as a function of TSD. a) 2D simulated I-V curve for three different TSD (or tunneling gap). b) as (a) but different scale showing that the I-V curve only marginally changes but $V_{OC}$ is independent of TSD c) Dark and PC seperated in simulation.*

Figure S5 depicts the 2D simulation of energy band diagram of the SHJ solar cell. Figure 5a shows the variation of conduction band, valance band, Fermi level of holes, and Fermi level of electrons in the vertical direction (z-axis) of the SHJ as indicated in Figure 1a of the main text at different applied bias. Figure S5b and S5c shows The shift of quasi Fermi level of holes and electrons as a function of applied bias at different vertical location of the sample, respectively. The quasi Fermi level of holes as a function of bias only change at the interface while it remains almost constant at the surface and mid point of (p) a-Si layer. However, the quasi Fermi level of electrons shifs almost linearly with applied voltage above 0.5 V applied voltages.

Figure S6a and S6b displays the 2D simulation of *I-V* curves for different TSD (see the Figure S6 above) with larger and smaller scale, respectively. The result shows that the open-circuit voltage is the same for different TSD but it is close to the measured average open-circuit voltage. However, this simulation does not show local variation of open-circuit voltage. In the simulation localized electronics states and quantum phenomenon were not taken in to consideration which implies that the simulation is more likely does not represent the real experiment conditions. Figure S6c shows the simulated I-V under illumination and dark conditions and their difference. Their difference which is the PC looks more or less voltage independence and it is similar to what is observed on the homogenous area (see Figure 4 in the main text).

## 6. Gap dependence open-circuit voltage of CPs

Figure S7 displays the histograms of the open-circuit voltage ($V_{OC}$) extracted from the measured *I-V* curves on CPs for TSD from 0.2 nm to 0.7 nm. These results show that at a given TSD there exists broad distributions of $V_{OC}$ and at the same time, the center of these distribution shifts towards smaller value with decreasing TSD. The center of the distribution shifts from 925 mV to 590 mV as the TSD varies from 0.7 nm to 0.2 nm, respectively. Previously, variation of open-circuit voltage from 0.2 V to 0.5 V at nano-scale has been reported on silicon nanowire solar cell using conductive-probe force microscopy[15]. Since, our measurements were carried out on nano-scale level using cAFM which is sensitive to many complicated scenarios, several possibilities can contribute to the observed variation. When TSD is very large, for example 0.7 nm, we measured $V_{OC}$ close to 1.2 V which is above the expected thermodynamic limit. The $V_{OC}$ distribution center shift with TSD and its broad distribution at a given TSD can be attributed to combination of some effects that is related to the property of the sample. At a given TSD, each CPs are the result of many percolation domains (paths) that involve valance band-tail states. As TSD becomes smaller, the volume of the sample that is in tunneling range and accessible to the tip will be bigger which can result in larger percolation domains that involves band-tail states in the volume of the emitter and intrinsic buffer a-Si:H layer. As a result of this, the local effective resistance of percolation domains (paths) will be reduced and lead to a shift of $V_{OC}$ to a lower value (see the center shift from bottom to the top plot of Figure S7). This effect, in turn is one cause of the observed $V_{OC}$ distribution, since CPs are not all identical and, thus, the induced shift will not be identical for different CPs either. In addition to a variation of local effective parallel resistance with TSD, the presence of positively or negatively charged state(s) [defect or band-tail state(s)] or dipoles[16–18] nearby or along charge percolation paths can induce a local change of electric potential landscape (positive or negative) which can monitor or control by turning off and on the local charge percolation paths as evidenced by the random telegraph noise due to trapping and de-trapping process or charging and discharging process. Other groups using scanning tunneling microscopy on the surface of silicon doped GaAs sample have shown controlled charge switching on individually addressable single donor state and a sharp jump of dark tunneling current due to the charge state of a donor state[19]. In their report, they have shown a shift of $I$-$V$ curve of approximately by 0.4 V for higher applied voltage to get the same dark tunneling current. This result is consistent with our observation of local $V_{OC}$ shift. Therefore, this effect can contribute to local $V_{OC}$ close to the band gap of c-Si (see 2D potential simulation below).

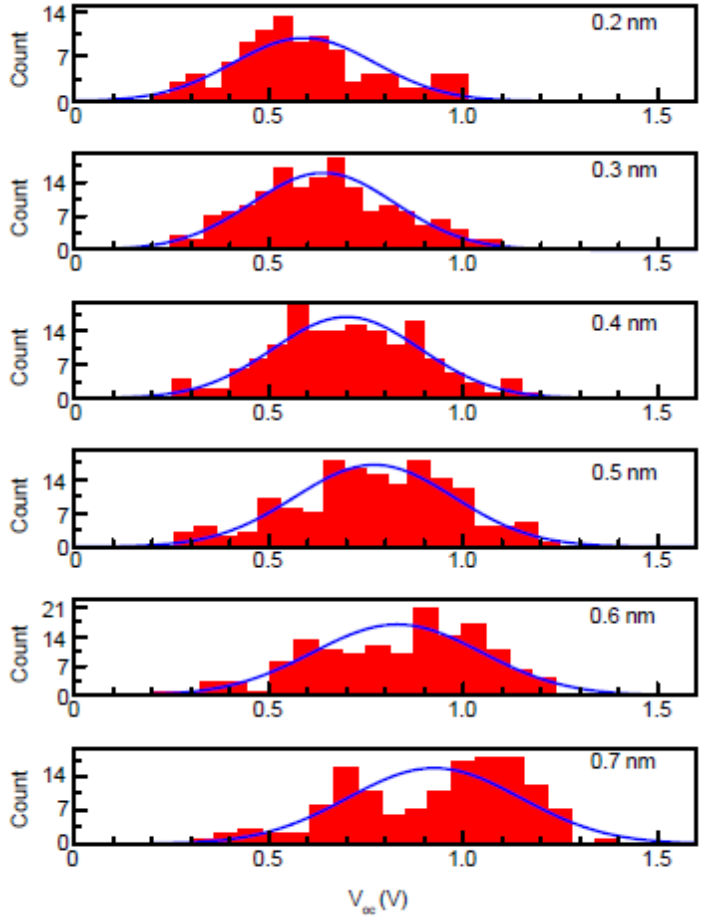

*Figure S7. Histogram displaying the variation of local open-circuit voltage determined with cAFM on CPs for six different TSD.*

**7. Simulation of local potential fluctuation.**

Figure S8 shows the 2D simulation of electrostatic surface potential due to the presence randomly distributed both negative and positive discrete point charges. Here we consider a simple electrostatic situation to check on the influence of discrete point charges on the surface potential at various TSD and hence the measured open-circuit voltage. For this simulation, we randomly distribute 20 discrete point charge (10 electrons and 10 holes) across the volume of a sample of 200 nm × 200 nm × 5 nm, representing our a-Si:H heterocontact. The total electrostatic surface potential at a distance $d$ away from the sample surface (representing TSD) was then calculated by summing the electric potential contribution of each point charges at each grid point above the sample surface. The potential was calculated for $128 \times 128$ grid points within the area of $100$ nm $\times$ $100$ nm. Figure S8a and S8b show the simulated electrostatic surface potential at distance $d = 0.2$ nm and $d = 0.7$ nm above the sample surface close to the experimental condition, respectively. Figure S8c shows the difference between surface potential of Figure S8a and S8b. Figure S8d, S7e and S8f shows the repetition of the simulation for another randomly distributed discrete point charges where in this case distribution (or locations) of point charges are different from the first case. These simulations results suggest that the presence of positively (negatively) charged trapped states can induce an increase (decrease) of local surface potential and hence contribute to the local $V_{OC}$ variations possibly close to the bandgap of c-Si. Similar phenomena of local surface potential variations have been observed for some systems such as insulator and semiconductor surfaces due to point charges located at the surface or subsurface [20,21], charging and discharging of quantum dots[22], charge trapping in high-dielectric amorphous and polycrystalline $Al_2O_3$ layers, and fixed charges in semiconductor heterojunction interface[23]. Particularly, Teichmann et al. have shown the mapping of single donor coulomb potential at the semiconductor-vacuum interface where its charging state controls the tunneling current[19]. Therefore, we attribute local fluctuation of open-circuit voltage to the presence of charged trap states in the HIT surface or subsurface or interface solar cell sample.

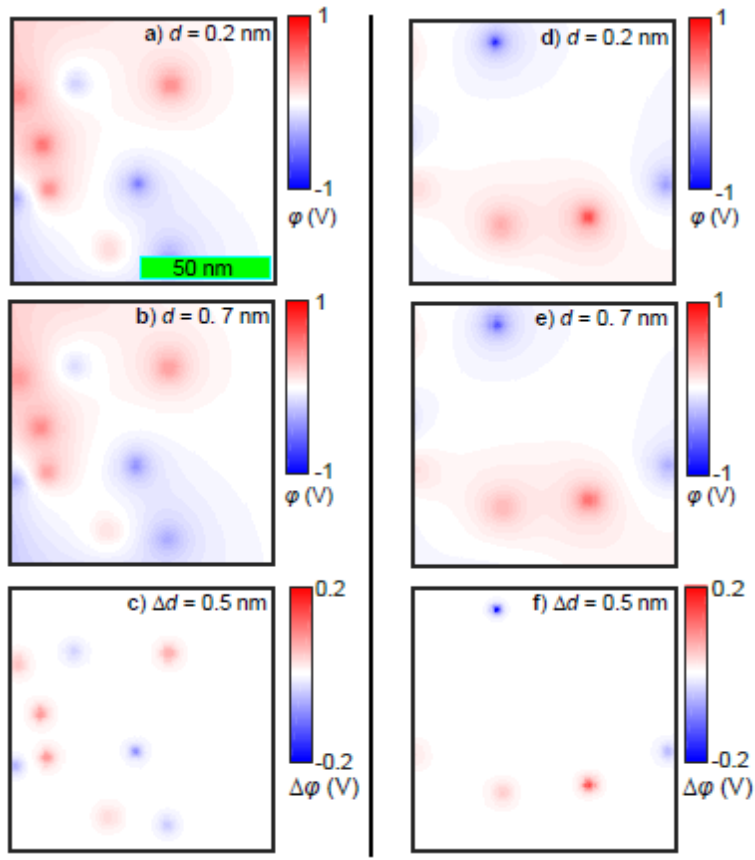

*Figure S8. Simulation of local potential fluctuation. a) Surface potential at a distance $d = 0.2$ nm. b) Surface potential at a distance $d = 0.7$ nm. c) Change in surface potential between distance $d = 0.2$ nm and $d = 0.7$ nm. d), e), and f) are surface potential at a distance $d = 0.2$ nm, $d = 0.7$ nm, and their difference, respectively.*

## 8. Exclusion of artifact signals for random telegraph noise (RTN) measurements

The RTN shown in Fig. 3e of the main text was, under the given conditions, qualitatively reproduced at about one quarter of the CPs where this experiment was repeated, yet this alone does not proof that this effect is not an artifact, e.g. due to weakly bond surface atoms which swtich their location randomly between the tip and the surface[38,39]. To test these observations for such artifacts, we identified first the a location of a pronounced CP which displayed RTN [see blue circled area in Figure S9a] and then a location within a HA were no RTN was seen [see red circled area in Figure S9a]. We then moved the tip between these two circled locations and we monitored both, the current and the frequency shift of the qPlus sensor [see Figure S9b through S9e]. While we observed no bimodal switching behavior for the frequency shift at any time, confirming that the RTN of the sample current is not caused by RTN of the TSD [see Figure S9c]. We also tested whether the RTN is highly reproducible at the blue circled location [see Figure S9b and S9e] while, at the same time, the absence of RTN at the red circled location was determined to be well reproducible, too. From these experiments, we conclude that the observed RTN must be caused by a mechanism that is inherent to the position where the RTN is detected, i.e. due to an electronic trap and, thus, we conclude that the sample current at a few CPs is switched between two levels by local charge trapping and reemission processes.

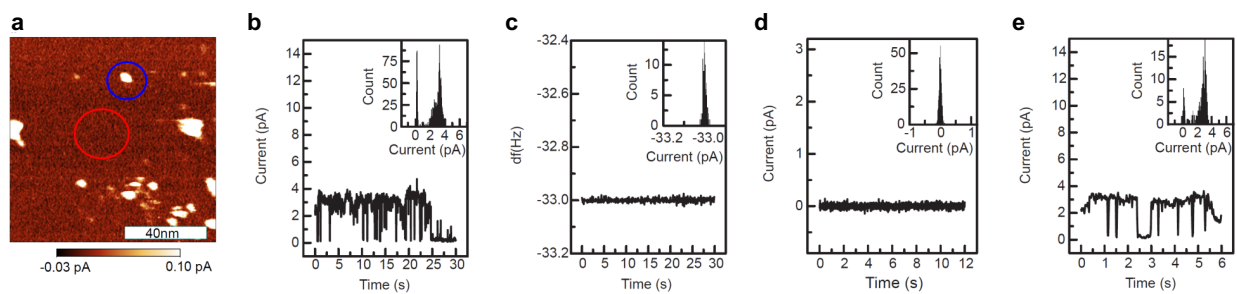

*Figure S9. (a) Image of CuI record at an arbitrary surface location. b) RTN measured on CP, circled blue. The inset is its histogram. c) Typical frequency shift as a function of time measured on CP and HA. The inset is the histogram. d) Current transient measured at location circled by the blue circle. The inset is the histogram. e) RTN measured at bright spot circled by the blue circle. RTN at bright spot shows a bimodal distribution while current transient off bright spot and frequency shift show unimodal distribution. See detail explanation in the text.*

# References

1.  Zhou, J. H. & Elliott, S. R. Tunneling recombination and the photoconductivity of amorphous silicon in the temperature region around 100 K. *Phys. Rev. B* **48**, 1505–1511 (1993).

2.  Tsang, C. & Street, R. A. Recombination in plasma-deposited amorphous Si:H. Luminescence decay. *Phys. Rev. B* **19**, 3027–3040 (1979).

3.  Street, R. A., Knights, J. C. & Biegelsen, D. K. Luminescence studies of plasma-deposited hydrogenated silicon. *Phys. Rev. B* **18**, 1880–1891 (1978).

4.  Wang, R., King, S. W. & Williams, C. C. Atomic scale trap state characterization by dynamic tunneling force microscopy. *Appl. Phys. Lett.* **105**, 052903 (2014).

5.  Wang, R. & Williams, C. C. Dynamic tunneling force microscopy for characterizing electronic trap states in non-conductive surfaces. *Rev. Sci. Instrum.* **86**, 93708 (2015).

6.  Zhang, Y. *et al.* Charge percolation pathways guided by defects in quantum dot solids. *Nano Lett* **15**, 3249–3253 (2015).

7.  Lanza, M., Iglesias, V., Porti, M., Nafria, M. & Aymerich, X. Polycrystallization effects on the nanoscale electrical properties of high-k dielectrics. *Nanoscale Res. Lett.* **6**, 108 (2011).

8.  Zhao, Z., Chen, X., Wu, H., Wu, X. & Cao, G. Probing the photovoltage and photocurrent in perovskite solar cells with nanoscale resolution. *Adv. Funct. Mater.* **26**, 3048–3058 (2016).

9.  Ambal, K. *et al.* Electrical current through individual pairs of phosphorus donor atoms and

silicon dangling bonds. *Sci. Rep.* **6**, (2016).

10. Abràmoff, M. D., Magalhães, P. J. & Ram, S. J. Image processing with ImageJ Part II. *Biophotonics Int.* **11**, 36–43 (2005).

11. Igathinathane, C. *et al.* Machine vision based particle size and size distribution determination of airborne dust particles of wood and bark pellets. *Powder Technol.* **196**, 202–212 (2009).

12. SYNOPSYS. *Sentaurus Structure Editor User Guide*. (synopsis, Inc, 2011).

13. Varache, R. *et al.* Investigation of selective junctions using a newly developed tunnel current model for solar cell applications. *Sol. Energy Mater. Sol. Cells* **141**, 14–23 (2015).

14. Varache, R., Kleider, J. P., Gueunier-Farret, M. E. & Korte, L. Silicon heterojunction solar cells: Optimization of emitter and contact properties from analytical calculation and numerical simulation. *Mater. Sci. Eng. B* **178**, 593–598 (2013).

15. Mikulik, D. *et al.* Conductive-probe atomic force microscopy as a characterization tool for nanowire-based solar cells. *Nano Energy* **41**, 566–572 (2017).

16. Wager, J. F. & Kuhn, K. Device physics modeling of surfaces and interfaces from an induced gap state perspective. *Crit. Rev. Solid State Mater. Sci.* **42**, 373–415 (2017).

17. Krüger, J., Bach, U. & Grätzel, M. Modification of TiO2 heterojunctions with benzoic acid derivatives in hybrid molecular solid-state devices. *Adv. Mater.* **12**, 447–451 (2000).

18. Reichel, C. *et al.* Electron-selective contacts via ultra-thin organic interface dipoles for silicon organic heterojunction solar cells. *J. Appl. Phys.* **123**, 24505 (2018).

19. Teichmann, K. *et al.* Controlled charge switching on a single donor with a scanning

tunneling microscope. *Phys. Rev. Lett.* **101**, 076103 (2008).

20. Boer, E. A., Bell, L. D., Brongersma, M. L., Atwater, H. A. & Watson, T. J. Models for quantitative charge imaging by atomic force microscopy. *J. Appl. Phys.* **90**, 2764–2772 (2001).

21. Huang, Z.-H. The image potential in scanning tunneling microscopy of semiconductor surfaces. *J. Vac. Sci. Technol. B* **9**, 2399–2404 (1991).

22. Salem, M. A., Mizuta, H. & Oda, S. Probing electron charging in nanocrystalline Si dots using Kelvin probe force microscopy. *Appl. Phys. Lett.* **85**, 3262–3264 (2004).

23. Kobayashi, Y. *et al.* Modulation of electrical potential and conductivity in an atomic-layer semiconductor heterojunction. *Sci. Rep.* **6**, (2016).